\documentstyle [12pt]{article}

\input epsf
\date{October 26, 1993}

\begin{document}

\begin{titlepage}

\nopagebreak

\vglue 1.5  true cm
\begin {center}
{\Large \bf
Wilson loop for large N Yang-Mills theory \\
\medskip
on a two-dimensional sphere }
\vglue 1.5 true cm
{\bf Jean-Marc DAUL }

\medskip
{\em Laboratoire de Physique Th\'eorique de
l'Ecole Normale Sup\'erieure\footnote{\rm Unit\'e Propre du
Centre National de la Recherche Scientifique,
associ\'ee \`a l'\'Ecole Normale Sup\'erieure et \`a
l'Universit\'e
de Paris-Sud.
24 rue Lhomond, 75231 Paris CEDEX 05, France.

\noindent
e-mail address : daul@physique.ens.fr, kazakov@physique.ens.fr}.}

\medskip
{\bf Vladimir A. KAZAKOV}

\medskip
{\em Department of Physics and Astronomy, Rutgers University, Piscataway, NJ
08854}

{and}

{\em Laboratoire de Physique Th\'eorique de
l'Ecole Normale Sup\'erieure}
\end {center}
\vfill
\begin{abstract}
\baselineskip .4 true cm
\noindent

We calculate various Wilson loop averages in a
pure $SU(N)$-gauge theory on a two-dimensional sphere, in the large $N$ limit.
The results can be expressed through the density of rows in the most probable
Young tableau found in \cite {dougkaz}.
They are valid in both phases (small and large areas of the sphere).
All averages for self-intersecting loops can be reproduced from the average for
a simple
(non self-intersecting) loop by means of loop equations.
{\footnotesize
}
\end{abstract}

\medskip
LPTENS 93/37

RU-93-50

October 1993
\vfill
\end{titlepage}
\def \tr { {\rm Tr}}
\def \wil {W(A_1,A_2)}
\def \ipi { {1\over 2 i \pi}}
\def \ippi { {1\over {(2 i \pi)^2}}}
\def \dd  {\partial}

\section{Simple non self-intersecting loop}

Two-dimensional Yang-Mills theory is quite a trivial example of a quantum
 field theory, since it has no propagating degrees of freedom.
 Nevertheless, the study of different gauge invariant quantities shows
 quite a rich structure and can be viewed as a useful tool for the
verification of mathematical hypotheses about the corresponding theory in
physically
 relevant dimension; say, for the search of a string theory describing
multicolour QCD.

As  was demonstrated in \cite{eqboucles,kostov} on the example
of Wilson loops, and
in \cite{grota} on the example of partition functions on the compact
two-dimensional manifolds, the results can be represented in terms of a sum
over minimal coverings (``soap films'') spanned on a contour, or wrapping over
the closed manifold. In this respect, the theory can have a beautiful
mathematical application, as was the case with the application of
three-dimensional Chern-Simons gauge theory.

First we shall consider a simple loop on a sphere,
separating areas $A_1$ and $A_2$,
with
which a holonomy $U$ is associated. The  average over all possible gauge
fields inside and outside the loop gives the partition function of
corresponding discs with fixed holonomies $U$ and $U^+$
 around the boundary, each
 given by the heat kernel (on the  group)
\begin{equation}
\big\langle 1|e^{-{A\over 2N} \Delta} |U \big\rangle =
\sum_{R} d_R e^{-{A \over 2N} c_R^{(2)}}
\end{equation}
 and
\begin {equation} \wil = \big\langle {1\over N} {\rm Tr} U \big\rangle
={1\over Z} \int dU {1\over N}  {\rm Tr} U \big\langle 1| e^{- {A_1 \over 2
N}\Delta} | U \big\rangle
 \big\langle U | e^{- {A_2 \over 2 N}\Delta} |1 \big\rangle \end {equation}
where $\Delta$ denotes the laplacian on $SU(N)$, with characters as
eigenfunctions:

\noindent
$(\Delta \chi _R)(U) = c^{(2)}_R \chi _R (U)$, $c^{(2)}_R$ being the quadratic
Casimir of
the irreducible representation $R$.

The character expansion of $\wil$ reads:
\begin {equation}
\wil = {1\over Z} \sum_{R_1,R_2} d_1 d_2 {1\over N} \int dU
\tr U \chi_1(U)
\overline{\chi_2(U)} e^{-{A_1\over 2N} c_1 + {A_2 \over 2N} c_2 }
\end {equation}
where $ \int dU \tr U \chi_1(U) \overline{\chi_2(U)}$ is the number of
occurrences of $R_2$
in $fundamental\ \otimes~R_1$, that is: $1$ when the Young diagram of $R_2$
is obtained by adding one box to the diagram of $R_1$ or $0$ otherwise.

Now, if the rows in the first diagram have lengths $(n_i)_{i=1\ldots N}$, the
corresponding dimension
is \begin {equation}
d=\prod_{1\leq i <j \leq N}{j-n_j -i +n_i \over j-i}
\end {equation}
and the Casimir
\begin {equation}
c= \sum_{i=1}^N n_i(n_i+N+1-2i)\ -N\, \langle n \rangle^2
\end {equation} with $\langle n \rangle ={1\over N}\sum_j n_j$.
If the $i^{th}$ row can be added
a box, the dimension is multiplied by \begin {equation} {d_2\over d_1} = \prod
_{j,j\neq i}
{j-n_j -i +n_i +1 \over j-n_j -i +n_i} \end {equation}
and the Casimir changes according to \begin {equation}c_2/N  = c_1/N + 2({n_i
-i \over N}+{1\over 2}-
{\langle n \rangle \over N}) \end {equation}

\noindent
We use the notation $\Phi _i =i-{N\over 2} - n_i +\langle n \rangle $ and $\phi
(x)$ its large N limit:
$ \phi (i/N)= \Phi _i /N $; we have to compute
\begin {equation}\wil = {1\over Z}\sum _{R_1} {1\over N}\sum_i d_1^2  \prod
_{j,j\neq i}
(1+{1\over \Phi_j -\Phi_i})
\ e^{-{A_1+A_2 \over 2 N } c_1 }\, e^{A_2 \phi(i/N)} \end {equation} where the
sum over $i$ corresponds to all
possible ways of adding one box to the diagram. This is the expectation value
\begin {equation}\big\langle {1\over N}\sum_i \prod _{j,j\neq i} (1+{1\over
\Phi_j -\Phi_i})\ e^{A_2 \phi(i/N)}
\big\rangle \end {equation} computed through averaging over all possible Young
diagrams with the same weights
as those which appear in the partition function of a sphere with total area
$A_1+A_2$; but in \cite {dougkaz}
has been found the shape of the most probable diagrams: if $x$ corresponds to a
line to which a box
can be added, we have $(A_1+A_2) \phi (x) = 2 \int_0^1 dy {1\over
\phi(x)-\phi(y) }$ and we can solve
for $\rho(\phi)={d\phi \over dx}$, obtaining a semi-circle law\footnote
{$\rho(\phi)={A\over 2\pi}\sqrt{{4\over A}-\phi^2}$} if $A=A_1+A_2 <
\pi^2$ or an expression
involving the complete elliptic integral of the third kind\footnote
{The even density is $\rho(\phi)={2\over \pi a \phi}
\sqrt{(a^2-\phi^2)(\phi^2-b^2)}
\Pi_1\big(-{b^2 \over \phi^2},k={b\over a}\big)$ when $b<\phi <a$ or $\rho =1$
when $-b<\phi<b$, with $aA=4K\ ,\  a(2E-k'^2K)=1$. }
if $A=A_1+A_2 > \pi^2$
(see \cite{dougkaz}).
Indeed, as the same weight is given to any representation and to its conjugate
(represented by
$n_i ' =2\langle n \rangle-n_{N+1-i}$), the unique solution for $\phi$ is
symmetric:
$\phi(1-x)=-\phi(x)$, and corresponds to (pseudo-)real representations.

In a similar way, we can average over $R_2$ and sum over all possible ways of
suppressing one box,
so that $\wil$ is also given by the average
\begin {equation} \big\langle {1\over N}\sum_i \prod _{j,j\neq i} (1+{1\over
\Phi_j -\Phi_i})^{-1}
\ e^{-A_1 \phi(i/N)}\big\rangle \end {equation}

We turn to the computation of \begin {equation}\exp \Big(\sum_{j,j\neq i} \ln
\big(1-{1\over \Phi_i -
\Phi_j} \big)\, \Big) \end {equation}
where the $\Phi_j$'s are distributed according to the above-mentioned law, with
fluctuations of order 1.
When $j$ is far apart from $i$ ($|j-i|$ of order $N$), the logarithm is of
order $1/N$, and there are
$N$ such terms: these give a finite contribution to the sum. But the same is
true for those $j$'s that
are close to $i$: there are a few such terms, each of order one, and one has to
take them into account,
adding their contribution to the (principal value) integral which accounts for
the terms of the first kind:
$-\int {dy \over \phi(x) - \phi(y)}$.
 Note that the latter terms were discarded in the computation of the shape
of the most probable diagrams: looking for a maximum of $d^2 \exp{\big(
-{Ac\over 2N}\big) }$
with its continuous formulation really means looking for a diagram with maximal
weight with respect to long-wavelength perturbations ($\delta n_i-\delta
n_{i+1}=0$ but
for few $i$'s, far apart from one another), in which case the contribution we
are going to
consider vanishes; while here it does not, because we consider a
short-wavelength
perturbation $\delta n_i= \delta_{i,i_0}$. So,
the solution of \cite {dougkaz} has to be understood in this way, and we have
to consider
the contributions of all the ``most probable'' diagrams, which differ from one
another
through short-wavelength fluctuations.

As the integers $\Phi_j$ have independent fluctuations of order 1, it seems
reasonable to
replace $\Phi_i - \Phi_j$
(for $j-i$ of order 1) by its continuous approximation $N
(\phi(i/N)-\phi(j/N))$, which is:
$(i-j) \phi'(i/N)$ in the large $N$ limit. Thus, the contribution
to the sum of logarithms is
\begin {equation}
-\sum_{j,j\neq i} \sum_{k\geq 1}{1\over k} (i-j)^{-k} \rho^k(\phi)
\label {eq.zeta}
\end {equation}
If we introduce $f(x)=\sum_{k\geq 1}{1\over 2 k} x^{2k} (B_{2k}/(2k)!)$ ($B_p=
p^{th}$ Bernoulli number),
and use the characterization of the most probable distribution, we obtain:
\begin {equation}\wil = {1\over N}\sum_i \exp \Big[ {A_2-A_1 \over 2}
\phi({i\over N}) + f(2i\pi
\rho(\phi))\Big] \end {equation}
Noting that $x f'(x)=x/2 -1 +x/(e^x-1)$, we have $f(x)=\ln \big({e^x-1 \over
x}\big) - {x\over 2}$ and
\begin {equation} \wil = {1\over N}\sum_i e^{\phi\ \delta A /2}{\sin(\pi
\rho)\over \pi \rho } \end {equation}
with $\delta A = A_2-A_1$; starting from an average over $R_2$, and considering
all possible ways of
suppressing one box, we would obtain the same expression with $A_2$ and $A_1$
interchanged: indeed,
the sign of $\delta A$ does not matter as the even distribution $\rho$ has only
even moments.

What about the sum over all acceptable lines?
Defining $n(x)$ to be $n_i /N$ ($x=i/N$), we see that if $|{dn\over dx}| >1$,
generically we have
$n_i>n_{i+1}$ and all lines are likely to be extended.
However, if $|{dn\over dx}| <1$, there is a finite fraction of lines for which
$n_i = n_{i+1}$, and
the non-increasing condition on the $n_i$'s prevents a finite fraction of lines
from being extended.
In this case, the contribution of $\displaystyle {1\over N}\sum_{acceptable \
i}$ in the range $dx$
is $dx |{dn\over dx}| = dx \big( {1\over \rho }-1 \big) $.

Finally, we would expect:
\begin {equation} \wil = \int d\phi {\rm Min}\big( {1\over \rho }-1,1
\big)e^{\delta A /2 \ \phi}
{\sin(\pi \rho)\over \pi}\end {equation}

However, as Boulatov first noticed {\cite {boulatov}}, one has to be careful in
expressing the product
$\prod _{j,j\neq i} (1+{1\over \Phi_j -\Phi_i}) $ in terms
of the continuous function $\phi$, when $|{dn\over dx}| <1$. In this case, he
groups factors in packets
and obtains a continuous approximation to $d_2/d_1$ which is the same as the
preceding one, but for
a factor $\rho \over 1-\rho$ that compensates for the ${\rm Min}(\ldots )$.
Indeed, when $|{dn\over dx}| <1$, fluctuations of lengths of the rows are
prevented by the non-increasing
condition on the $n_i$'s: this is the very reason why a direct continuous
approximation is not valid.
We shall rather consider the heights of the columns, which are free to
fluctuate, so that the argument
given above for lines gets transposed.

\noindent
When $|{dn\over dx}| <1$, generically we have : $n_i-n_{i+1}=0 \  {\rm or} \
1$, so columns are labelled
by $n$; summing over all possible additions of one box means summing over all
$n$'s.
Just as in the preceding case, the change in the Casimir is ${c_2-c_1\over
N}=-2\phi$; as for dimensions,
the contribution of $j$'s far apart from $i$ reduces to the exponential of the
same principal value integral,
recognized as $\exp \big(-{A_1+A_2\over 2} \phi \big) $. To study the
contribution of $j$'s close to $i$
we consider a truncated diagram, formed with a finite number of lines extending
above and under $i$; let
$h_n$ denote the height of the $n^{th}$ truncated column ($\nu_1 \leq n \leq
\nu_2 $) and set
$\tilde \Phi_n =n-h_n\ ,\ \tilde \phi = \tilde \Phi / N$. How does the
dimension change when we add a box
at the end of line $i$, that is at the bottom of column $n=n_i+1$ ?

\noindent
Before the extension, $\prod_{j<i} ( \Phi_i- \Phi_j)$ contains all integers
ranging from $1$ to
$h_n+\nu_2-n+1$ but the $h_n-h_{\nu}+\nu -n+1\ ,\nu=n+1,\ldots,\nu_2$.
Similarly, the product
$\prod_{j>i} (\Phi_j-\Phi_i)$ evaluates to $\displaystyle (\tilde \Phi_n
-\tilde \Phi_{\nu_1}-2)! \over
{\displaystyle \prod_{\nu_1 <\nu <n}(\tilde \Phi_n -\tilde \Phi_\nu -1)} $.

\noindent
After the extension these factors are respectively
$\displaystyle (h_n+\nu_2-n)!\over {\displaystyle \prod_{n < \nu \leq \nu_2
}(\tilde \Phi_\nu -\tilde \Phi_n)}
$ and $\displaystyle (\tilde \Phi_n -\tilde \Phi_{\nu_1}-1)! \over
{\displaystyle \prod_{\nu_1 <\nu <n}(\tilde
\Phi_n -\tilde \Phi_\nu)} $,
so that:
\begin {equation}{\tilde d_2 \over \tilde d_1}= \prod_{\stackrel{\nu_1 <\nu
\leq \nu_2}{\nu \neq n}}
\Big(1+{1\over \tilde \Phi_\nu -\tilde \Phi_n} \Big) \ \ \times \ {\tilde
\Phi_n -\tilde \Phi_{\nu_1}-1 \over
\nu_2 - \tilde \Phi_n +1} \end {equation}
(we have just expressed the (change in) dimension in terms of heights of
columns for ``steep'' diagrams:
$n_i - n_{i+1} = 0\ {\rm or}\ 1$).

\noindent
The last factor goes to 1 for large truncated diagrams, extending {\em
symmetrically} around $i$
(use $h_p -h_n = {dh\over d\nu}(p-n)$ for large $N$, and note that this
symmetry corresponds to
the cancellation of $\zeta (2p+1)$ in (\ref {eq.zeta}) and to the
identification of the other terms
with a principal value integral).

Finally, arguing that the heights of columns are integers with fluctuations of
order 1, and can be
approximated by a continuous function, we apply the same computation as before
to get:
\begin {equation}\wil = {1\over N}\sum_n  e^{\delta A /2 \ \phi}\ {\sin(\pi
\tilde \rho)\over
\pi \tilde \rho }\end {equation}
where $\tilde \rho = {dn\over d\tilde \phi}=-{dn \over d\phi }= 1-\rho $.
This is exactly:
\begin {equation}
 \wil = \int d\phi e^{\delta A /2 \ \phi}\, {\sin(\pi \rho)\over \pi}
 \label {eqgale}
\end {equation}

The same result was independently obtained by Boulatov \cite {boulatov}.

\section {Other loops}

Before showing how the loop equations \cite {migdalmak,eqboucles} allow us to
compute
self-intersecting
Wilson loops, we cast (\ref {eqgale}) into a more convenient form for later
purposes.

Define $G(z)=\int d\phi{\displaystyle \rho(\phi) \over z-\phi}$: if $x$ lies in
the support of $\rho$ and is such
that $\rho (x)<1$ ($x$ corresponds to a region where boxes can be added) we
have
\begin {equation}
G(x\pm i\epsilon)={A\over 2} x \mp i \pi \rho(x)
\end {equation}
with $A$ the total area; if $\rho(x)=1$ then $\sin(\pi \rho)$ vanishes.
Thus, we can write
\begin {equation}
\wil = \ipi \oint dz\, e^{G(z)-A_1 z} = \ipi \oint dz\, e^{G(z)-A_2 z}
\end {equation}
where the contour of integration goes around the cut of $G$ in the positive
direction.

Other loops can be easily obtained by the methods of \cite {eqboucles} and
expressed as contour
integrals. Consider for instance the following case:
\vskip 5mm
\parbox {1pt}
{\epsfysize=100pt \epsfbox{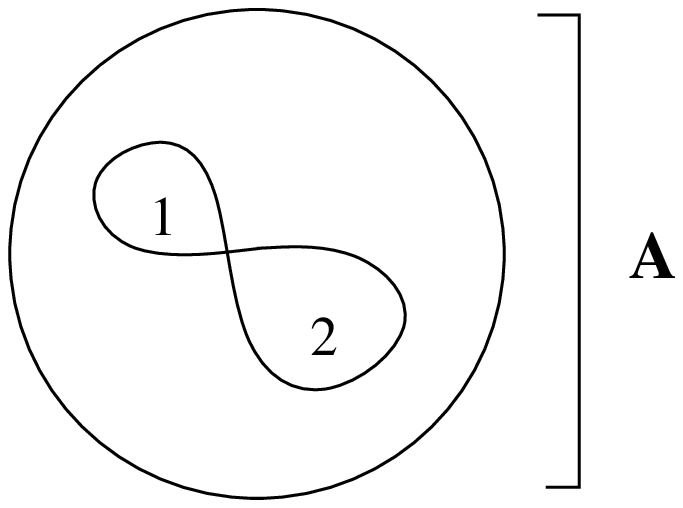}}
\parbox{300pt}
{\em \hspace {10mm} An 8-like contour on the sphere}
\vskip 5mm
The associated Wilson average $W_{1,2}$ satisfies $-(\partial_1
+\partial_2)W_{1,2} = W_1 W_2$, where
we note $W_i$ for a simple loop enclosing area $i$ and $\partial_i$ for
derivation with respect to
area $i$ (the total area $A$ being kept fixed). We now integrate this equation
to obtain :
\begin {equation}
W_{1,2}=\ippi \oint dx\, dy\, {e^{G(x)+G(y)-A_1 x-A_2 y}\over x+y}
\label {huit}
\end {equation}
Indeed, we have just seen that both sides vary by the same quantity when the
areas 1 and 2 are changed
by the same amount; so, we only have to check that they coincide when one of
the areas vanishes.
Suppose $A_2=0$: the left-hand side of (\ref {huit}) reduces to $W_1$, and in
the right-hand side,
we use $G(y)\sim {1\over y}$ for $y\to \infty$ to compute the residue at
infinity of the
$y$ integral, and find $W_1$ again.

So,(\ref {huit}) holds provided the smallest area is associated with the outer
contour. But contours
can be interchanged: the difference between the integral with $y$ on the outer
contour and
the one with $y$ on the inner contour is given by the (sum of) residue(s) at
$-x$ (for each
value of $x$), that is
\begin {equation}
\ipi \oint dx \, e^{G(x)+G(-x)-(A_1-A_2)x}
\end {equation}
that is 0, because $G$ is an odd function. We conclude that contours can be
interchanged: their relative
position does not matter, provided they do not intersect (more properly: $x+y$
shall not vanish; otherwise,
the double integral is still convergent, but it does not give the right
answer).

As another example, we consider the Pochhammer contour
\vskip 5mm
\epsfysize=100pt \epsfbox{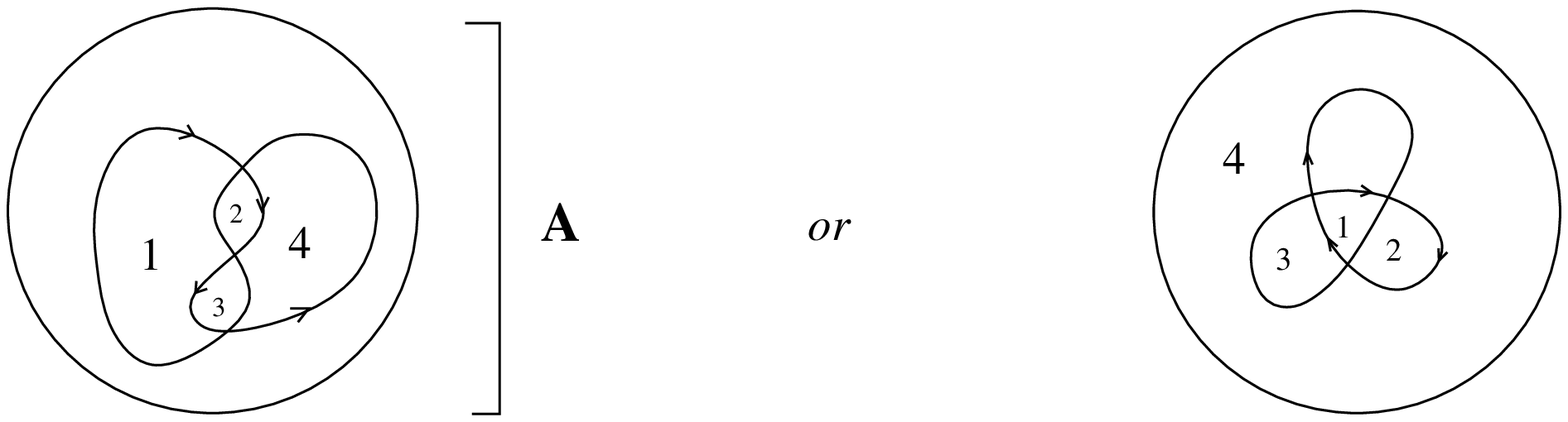}
\vskip 5mm
for which the loop equations give:
\begin {eqnarray}
 (\dd_3-\dd_4-\dd_1)W_{1,2,3,4} &= &W_{1+2} W_{2+4} \nonumber \\
( -\dd_4+\dd_2-\dd_1)W_{1,2,3,4} &= &W_{1+3} W_{3+4} \\
 ( \dd_2-\dd_4+\dd_3-\dd_1)W_{1,2,3,4} &= &W_{1+2+3} W_{2+3+4} \nonumber
\end {eqnarray}
so that $\dd_3W_{1,2,3,4} = W_{1+2+3} W_{2+3+4} - W_{1+3} W_{3+4}$ and
\begin {equation}
W_{1,2,3,4} =\ippi \oint dx\, dy\, e^{G(x)+G(y)-A_1
x-A_4y-A_3(x+y)}{1-e^{-A_2(x+y)}\over x+y}
\ + F_{1,2,4}
\end {equation}
We take $A_3=0$ to compute $F_{1,2,4}$: in this case the Pochhammer contour
reduces to an 8-like loop
($W_{1,4}$), and
\begin {equation}
F_{1,2,4} =\ippi \oint dx\, dy\, e^{G(x)+G(y)-A_1 x-A_4y}\, {e^{-A_2(x+y)}\over
x+y}
\end {equation}

Finally, we have:
\begin {equation}
W_{1,2,3,4} =\ippi \oint dx\, dy\, { e^{G(x)+G(y)-A_1 x-A_4y}\over x+y}\,
\Big(1-\big(1-e^{-A_3(x+y)}\big)
\big(1-e^{-A_2(x+y)}\big) \Big)
\end {equation}
(here again, the relative position of contours is arbitrary: we only require
$x+y$ not to vanish)
or $W_{1,2,3,4} =W_{3+1,3+4} +W_{2+1,2+4} - W_{2+3+1,2+3+4}$.

\section {The small area phase}

 When the total area $A=A_1 + A_2$ is less than $\pi^2$, the density $\rho$
obeys a semi-circle law
\cite {dougkaz,ptitaire}:
\begin {equation}\rho (\phi) = {A\over 2 \pi} \sqrt {{4\over A}-\phi^2} \end
{equation}
so that (\ref {eqgale}) gives:
\begin {equation}\wil = {2\over \pi}\int_0^{2/\sqrt A} d\phi \, \cosh \Big(
{\delta A \over 2} \phi \Big)
\, \sin \Bigg( {A\over 2  } \sqrt {{4\over A}-\phi^2} \Bigg) = \sqrt{A\over A_1
A_2} J_1 \Bigg(
2 \sqrt {A_1 A_2 \over A} \Bigg)
\label {bessel} \end {equation}

This result had already been obtained in \cite {dougkaz} by a direct estimation
of the heat kernel
\cite{chaleur},
with windings neglected:
\begin {equation} \wil \sim \int d\theta_{1\ldots N} \prod_{i<j} (\theta_i
-\theta_j)^2\, e^{-{N\over 2}
({1\over A_1 }+{1\over A_2})\sum \theta_i^2 } \ {1\over N}\sum_i \cos
(\theta_i) \end {equation}

At large $N$, the angles $\theta$ are distributed according to:
\begin {equation}\rho (\theta) = {1\over 2 \pi} \sqrt {A\over A_1 A_2}\ \sqrt
{{4 A_1 A_2\over A}-
\theta^2}\end {equation}
and finally:
\begin {equation}\wil = \langle \cos (\theta) \rangle = (\ref {bessel}) \end
{equation}

Apparently, windings (that is, periodicity for $\theta$'s) can be neglected
consistently provided
the support of $\rho$ doesn't extend out of $[-\pi,\pi]$, which is the case as
long as
${1\over A_1}+{1\over A_2}>{4\over \pi^2}$; in particular, one of the areas can
be arbitrarily large.
However, this is not the case with the derivation here given: the total area
has to be less than
$\pi^2$ for the semi-circle law to be valid; and, for one large area, Boulatov
\cite {boulatov}
checked the exponential law: indeed, he could reproduce the large-area
expansion given by
Gross and Taylor's prescriptions \cite {grota}.

\noindent
A deeper understanding of the contribution of windings seems needed to directly
handle the
heat kernel in these computations.

\bigskip
{\bf Acknowledgements}

We are grateful to M.Douglas for the fruitful discussions at the early stage
of this work, and to D.Boulatov for the exchange of results prior
 to publication.

\end{document}